\documentclass{baltzer}
\input {epsfig.sty}
\begin{document}
\begin{frontmatter}
%
\title{High-Precision Storage Ring for g-2 of the Muon and 
Possible Applications in Particle and Heavy Ion Physics
\thanks{This article describes in part results from work performed by the
muon g-2 collaboration at BNL and working groups on the muon neutrino mass and the
muon electric dipole moment.}}       
\author[A]{Klaus P. Jungmann }     
\address[A]{Physikalisches Institut, University of Heidelberg, 
D-69120 Heidelberg, Germany \email{jungmann@physi.uni-heidelberg.de}}   
\runningauthor{K. Jungmann}
\runningtitle{High-Precision Storage Ring for ...}
%
\begin{abstract}  
A new superferric magnetic storage ring with highly homogeneous field at
1.45~T and weak electrostatic focussing is described which has been set up at the 
Brookhaven National Laboratory (BNL), USA, for a precision measurement
of the magnetic anomaly of the muon. The toroidal storage volume has a radius
of 7~m and a diameter of 9~cm. Precision magnetic field determination 
based on pulsed NMR on protons in H$_2$O yields the field 
to better than 0.1~ppm everywhere within the storage region. Follow on 
experiments using the setup have been already suggested to search for a 
finite mass of the muon neutrino and to search for an electric dipole 
moment of the muon with significantly increased accuracy.
The high homogeneity of the field suggests the usage of such devices 
as a mass spectrometer for heavier particles as well.    
\end{abstract}
%
\end{frontmatter}
\section{Introduction}
Trapping of elementary particles in combined magnetic and electric fields
has been very successfully applied for obtaining properties of 
the respective species and for determining most accurate values of fundamental constants.
The magnetic anomaly of fermions $a=\frac{1}{2}\cdot(g-2)$
describes the deviation of their magnetic g-factor
from the value 2 predicted in the Dirac theory. It
could be determined for electrons and positrons in Penning traps
by Dehmelt and his coworkers to 10~ppb \cite{Dyc_90}. 
Accurate calculations involving
almost exclusively the "pure" Quantum Electrodynamics (QED)
of electron, positron and photon fields allow the most precise
determination of the fine structure constant $\alpha$ \cite{Kin_95}
by comparing experiment and theory 
in which $\alpha$ appears as an expansion coefficient.
The high accuracy to which calculations in the framework of QED can
be performed is demonstrated by the satisfactory
agreement between this value of $\alpha$ and the ones obtained in
measurements based on the quantum Hall effect \cite{QH} as well as
the ac-Josephson effect and the gyromagnetic ratio of protons in water
\cite{ACJ},
or the number extracted from the very precisely known Rydberg constant \cite{Ryd}
using an accurate determination of the neutron de Broglie wavelength \cite{NDBW}
and well known relevant mass ratios \cite{PDG}.
Moreover, the excellent agreement of $\alpha$ values determined 
from the electron magnetic anomaly 
and from the hyperfine splitting in the muonium atom \cite{Bos_96}
may be interpreted as the most precise 
reassurance of the internal consistency of QED, 
as the first case involves  QED of 
free particles  whereas in the second case distinctively 
different bound state QED approaches need to be applied \cite{Kin_90}. \\

The anomalous magnetic moment of the muon $a_{\mu}$ has a 
$(m_{\mu}/m_e)^2 \approx 4 \cdot 10^{4}$ times
higher sensitivity to heavier particles and other than electromagnetic interactions.
They can be investigated carefully, as very high confidence in the
validity of calculations of the dominating QED contribution arises from the
success of QED describing this quantity for the electron. 

For the muon  $a_{\mu}$ has been measured in a series of three 
experiments at CERN \cite{Far_90} all using magnetic muon storage.
Contributions arising from strong interaction amount to 60~ppm
and could be identified in the last of these measurements.    \\  

\begin{table}[thb]
\protect  \caption{
         Sensitivity to new physics of the g-2 experiment at BNL aiming 
         for 0.35~ppm relative accuracy.
         }
\vspace{2mm}
\protect \label{g2_new_physics}
{
  \begin{tabular}{lllll}
\hline
new physics     &\multicolumn{3}{c}{sensitivity}   &other experiments\\
\hline
\hline\\
Muon substructure    & $\Lambda$& $\geq$ & 5~TeV
& LHC similar\\
excited muon   & $m_{\mu^*}$& $\geq$ & 400~GeV
& LEP II similar\\
W$^\pm$-boson substructure & $\Lambda$& $\geq$ & 400~GeV
& LEP II $\sim$100-200~GeV \\
W$^\pm$ anomalous &$a_W$& $\geq$   & 0.02
& LEP II $\sim$0.05, \\
magnetic moment&&&& LHC $\sim$0.2\\
Supersymmetry                 & $m_{\widetilde {W}}$& $\leq$   & 130~GeV
& Fermilab p${\overline{\rm p}}$ similar\\
right handed W$^\pm_R$-bosons &$m_{W'}$& $\leq$    & 250~GeV
& Fermilab p${\overline{\rm p}}$ similar\\
%
%
heavy Higgs boson            & $m_H$& $\leq$        & 500~GeV
&       \\
Muon electric dipole moment & $D_{\mu}$& $\leq$ & $4\cdot10^{-20}ecm$
                                        &       \\
\hline
\end{tabular}
} 

 {\footnotesize
$^a$ for substructure $\Delta a_{\mu} \sim m^2_{\mu}/\Lambda^2$
}
\end{table}

At the Brookhaven National Laboratory a new dedicated experiment is being
set up to determine the muon's magnetic anomaly which 
aims for 0.35 ppm relative accuracy meaning a 20 fold improvement over 
previous results. 
At this level it will be particularly sensitive to contributions 
arising from weak interaction through
loop diagrams involving W and Z bosons (1.3~ppm). The experiment
promises further a clean test of renormalization in weak interaction.
The muon magnetic anomaly may  also contain
contributions from new physics \cite{Mer_90,Lop_94,Ren_97}.
A variety of speculative theories can be tested which try to
extend the present Standard Model in order to explain some of its
not yet understood features. This includes muon substructure, new gauge bosons,
supersymmetry, an anomalous magnetic moment of the W boson and leptoquarks.
Here this measurement is complementary 
to searches carried out in the framework of other high energy experiments.
In some cases the sensitivity is even higher (Table 1).

\section{The Brookhaven g-2 Magnet}

In the BNL experiment polarized muons are stored 
in a magnetic storage ring
of highly homogeneous field $B$ and with weak electrostatic focussing using
quadrupole electrodes around the storage volume.
The difference frequency
of the spin precession and the 
cyclotron frequencies,
$\omega_a = a_{\mu} \frac{e}{m_{\mu}c}B  $,
is measured,
with $m_{\mu}$ the muon mass and $c$ the speed of light,
by observing electrons/positrons from the weak decay
$\mu^{\pm}\rightarrow e^{\pm} + 2\nu$.
For relativistic muons the influence of a static electric field vanishes 
\cite{Telegdi}, if $ a_{\mu} = 1/( \gamma_{\mu}^2-1)$ 
which corresponds to $\gamma_{\mu}=29.3$ and a muon momentum of 3.094 GeV/c,
where $\gamma_{\mu}= 1/ \sqrt{1-(v_{\mu}/c)^2}$ and $v_{\mu}$ is
the muon velocity.
The momentum needs to be met at the 10$^{-4}$ level
for a corresponding correction to be below the desired accuracy for $a_{\mu}$.
For a homogeneous field the magnet must have iron flux return and shielding.
To meet this, the particular momentum requirement and 
to avoid magnetic saturation of the iron a device of 
7~m radius was built. It has a C-shaped iron yoke cross section with 
the open side facing towards the center of the ring.
It provides 1.4513~T field  in a 18~cm gap. The magnet is energized by 4
superconducting coils carrying 5177~A current. 
The storage volume inside of a Al vacuum tank has 9~cm diameter.

The magnetic field is measured by a newly developed 
narrow band magnetometer system
which is based on pulsed nuclear magnetic resonance (NMR) of protons in water.
It has the capability to measure the field absolute to $\approx$50~ppb
\cite{Pri_96}. The field and its homogeneity are  continuously monitored by 
366 NMR probes which are embedded in the Al vacuum tank and 
distributed around the 
ring. Inside the storage volume a trolley carrying 17 NMR probes arranged to
measure the dipole field and several important multipole components as well
a fully computerized magnetometer built from all nonferromagnetic components
is used to map the field at regular intervals. 
The accuracy is derived from and related
to a precision measurement of the proton gyromagnetic ratio in 
a spherical water sample  \cite{Phillips_77}.
The field homogeneity at present is about 25~ppm. It will be improved to the 
ppm level using mechanical shimming methods and a set of electrical shim coils.
The field integral in the storage region is known at present to better 
than 1 ppm at any time and field drifts of a few ppm/hour were observed.
It can be expected that with additional shimming and thermal insulation
for the magnet yoke the field path integral can be known to 0.1~ppm. 

The weak focussing is provided by  
electrostatic quadrupole field electrodes with 10~cm 
separation between opposite plates. 
They cover four $39^{\circ}$ sections of the ring.
The electric field is applied by pulsing
a voltage of $\pm 24.5$~kV for $\approx$ms  duration to avoid trapping of
electrons and electrical breakdown \cite{Flegel_73}.

Due to parity violation
in the weak muon decay process the positrons are emitted preferentially
in/opposite to the muon spin direction causing a time dependent
variation of the spatial distribution of decay particles in the muon 
eigensystem which translates into a time dependent variation of the energy
distribution observed by detectors fixed inside the ring. 
\begin{figure}[t]
\label{Fig1}
 \begin{minipage}{2.5in}
  \centering{
   \hspace*{-0.0in}
   \mbox{
   \epsfig{figure=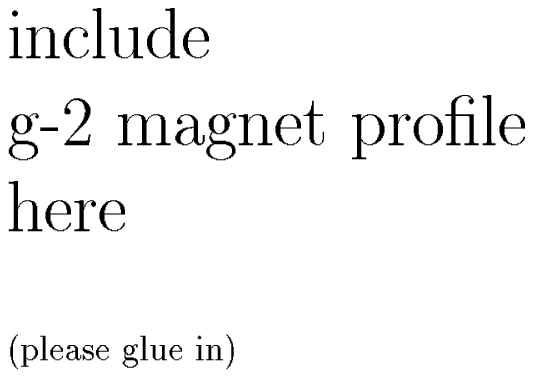,width=2.5in}
         }
             }
 \end{minipage}
 \hspace*{0.1cm}
 \begin{minipage}{2.5in}
  \centering{
   \hspace*{-0.0in}
   \mbox{
   \epsfig{figure=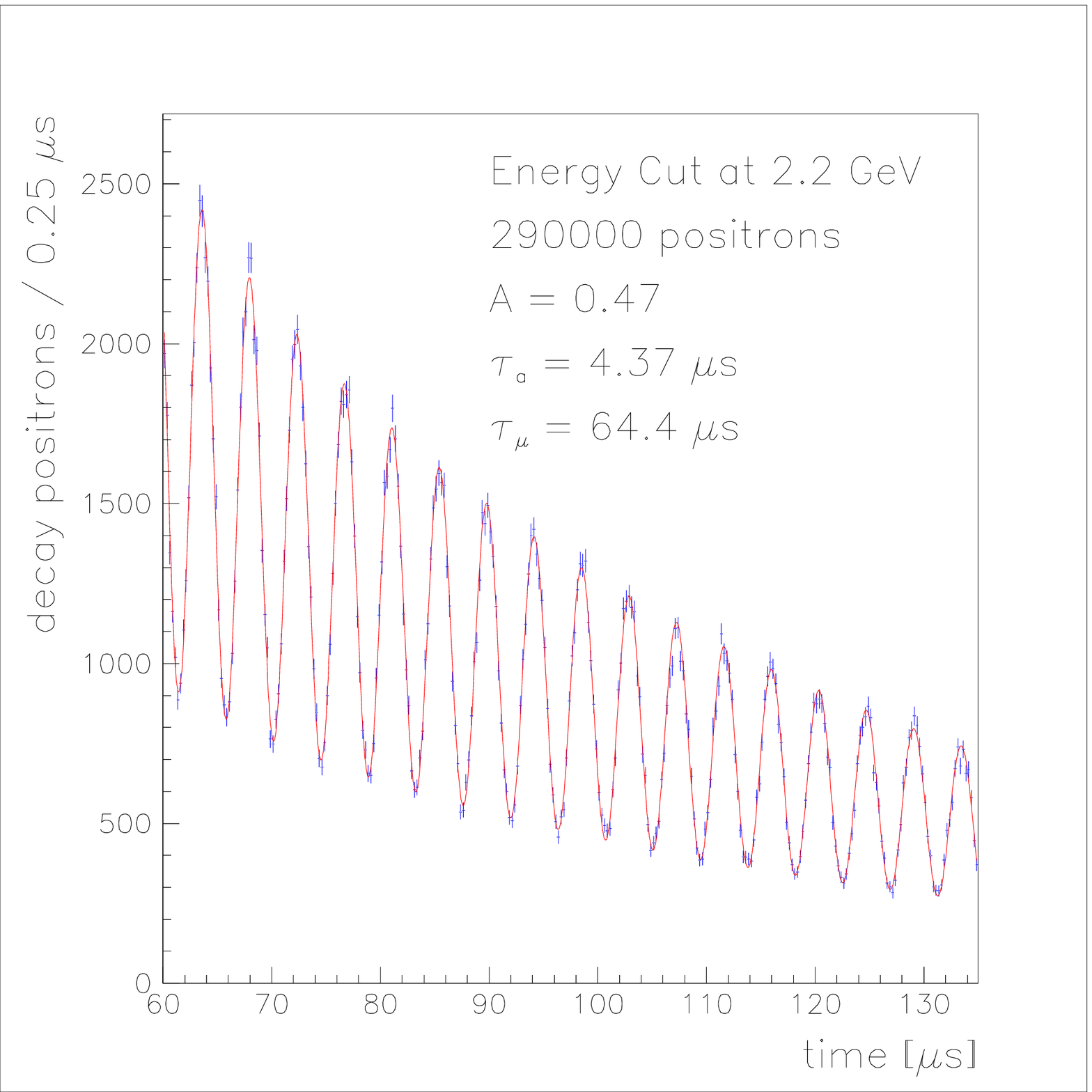,width=2.5in}
         }
             }
   \end{minipage}
 \centering\caption[]
        {Cross sectional view of the magnet iron construction (left).
         Preliminary analysis of the observed muon spin precession 
         signal in the new experiment (right).
        }

\end{figure} 

The improvements over previous experiments include
an azimuthally symmetric iron construction for the magnet 
with superconducting coils,
a larger gap and higher homogeneity of the field,
an electron/positron detector system
covering a larger solid angle and using segmented detectors
and improved electronics.
A major advantage is the two orders of magnitude  
higher primary proton intensity
available at the AGS Booster at BNL. A new feature will be ultimately
the direct injection of muons into the storage volume using an 
electromagnetic kicker
as compared to filling the ring
with decay muons from injected pions which has been employed so far. 

In order for the new muon g-2 experiment to reach its design 
accuracy 
besides the field also the muon mass respectively its magnetic moment 
needs to be
known to 0.1~ppm or better. An improvement beyond the present 0.36~ppm 
accuracy of this 
constant can be expected from both microwave spectroscopy 
of the muonium atom's ($\mu^+e^-$)
hyperfine structure and from laser spectroscopy of the muonium 1s-2s 
transition \cite{Jun_94}.\\ 

\section{Experiments beyond Muon g-2 using the Ring}

The large installation with the highly homogeneous field  seems to be 
attractive for further experiments. Of particular interest are
applications where the homogeneous field is used in a mass spectrometer.

\subsection{Particle Physics Experiments}

Among the seriously discussed suggestions 
are the use of the magnet for searching for a finite mass of the 
muon neutrino \cite{Prisca_97}. 
By injecting relativistic pions and comparing the positions of the 
decay muons after one single turn in the magnet with the ones of pions
one can expect a sensitivity to about 8~keV/c$^2$ 
which is 20 times better than the present limit \cite{Borch_97}. Due to 
relativistic kinematics this method
is less sensitive to the precision of the present knowledge of the pion mass compared to 
the previously employed technique \cite{Borch_97}.  \\

A very promising approach seems to be the proposed search for a permanent 
electric dipole moment of the muon \cite{Yannis_97}. 
For such an experiment major modifications of the magnet setup
would be required which involve the application of a radial electric field
and the switching from electrostatic to alternating gradient focussing by
replacing the pole tips of the magnet with appropriately shaped pieces of iron.
In case of a finite electric dipole moment a time dependent asymmetry in 
the muon decay rates
counted above and below the storage region is expected as a signature.
Such an experiment may achieve up to four orders of magnitude improvement
and could  reach a level of sensitivity at which several
theoretical models, particularly such involving supersymmetry, could be tested.

\subsection{Heavy Ion Physics Possibilities}

In the field of stored heavy ions unfortunately  
the weak focussing device cannot be expected to contribute to studies of and 
searches for 
crystalization of ion beams, as this phenomenon is prohibited by instabilities
caused by radial deplacement dependent particle precession \cite{Ruggiero}. 
However,
if the field homogeneity were sacrificed for alternating gradients and just the
basic circular topology of the iron yoke and the field exciting coils were
kept, one would have a chance to observe the effect.

The highly homogeneous field of the g-2 magnet looks promising
for precision mass measurements of heavy ions. The field homogeneity and the
accurate knowledge of the field integral can only be maintained, if the field 
is kept at a constant level. Therefore, only such experiments can be considered
which do not need any variation of the field, 
i.e. the momentum of the particles to be injected into the ring needs to be
adjusted appropriately. Among the possibilities one can expect 
accurate determinations
of differences between very close masses, e.g. isomers, 
through their slightly different orbits and
rotation frequencies. 
Neighbouring isotopes could be compared by additionally varying the
momentum of different particles to be compared in their masses 
prior to injection.

\section{Conclusions}

The new Brookhaven g-2 magnet is a powerful device which is expected to
guarantee the success of the experiment it has been designed for. Important
questions in particle physics can be addressed with minor modifications
subsequently. Finally the  concept of a circular uniform iron ring yoke 
with four circular coils for field excitation 
may find applications in precision heavy ion measurements.

\section{Acknowledgments}

The author would like to thank the organizers of the 
3$^{rd}$ Euroconference on "Atomic Physics with Stored Highly Charged Ions"
for the stimulating atmosphere in  nice surroundings and for
support. This work was sponsored in part by the German Bundesminister f\"ur 
Bildung und Forschung.
\\

{\footnotesize
The people who contributed within the g-2 collaboration to the present 
status of the experiment:
D.H. Brown, R.M. Carey, W. Earle, E. Efstathiadis,
E.S. Hazen,  F. Krienen,
J.P. Miller, V. Monich, J. Ouyang, O. Rind, B.L. Roberts$^{\ddag}$, 
L.R. Sulak, A. Trofimov, G. Varner,
W.A. Worstell {\bf Boston University};
 J. Benante, H.N. Brown, G. Bunce, J. Cullen, G.T. 
Danby, J. Geller,
H. Hseuh, J.W. Jackson, L. Jia, S. Kochis, R. Larsen, Y.Y. 
Lee,
W. Meng, W.M. Morse$^{\ddag}$, 
 C. Pai, I. Polk, R. Prigl,
S. Rankowitz, J. Sandberg,
Y.K. Semertzidis, R. Shutt, L. Snydstrup, A. Soukas, A. 
Stillman,
T. Tallerico, M. Tanaka, F. Toldo, D. Von Lintig, K. Woodle,
{\bf Brookhaven Nat. Lab.};
T. Kinoshita, Y. Orlov,
{\bf Cornell Univ.};
D. Winn,
{\bf Fairfield Univ.};
J. Gerhaeuser, W. Frauenfeld, 
A. Grossmann, K. Jungmann, G. zu Putlitz, P. von Walter,
{\bf Univ. of Heidelberg};
P.T. Debevec, W. Deninger,
D.W. Hertzog, C. Polly, S. Sedykh, D. Urner
{\bf Univ. of Ilinois};
U. Haeberlen,
{\bf MPI f.Med.Forschung, Heidelberg};
 P. Cushman, L. Duong, S. Giron, J. Kindem, R. 
McNabb,
D. Miller, C. Timmermans, D. Zimmerman {\bf Univ. of Minnesota};
L.M. Barkov, A. Chertovskikh, V.P. Druzhinin, G.V. 
Fedotovich, D.N. Grigorev,
V.B. Golubev, B.I. Khazin,  I. Logashenko, A. Maksimov,
Yu. Merzliakov, N.M. Ryskulov,
S. Serednyakov, Yu.M. Shatunov, E. Solodov, 
{\bf Inst. of Nucl. Phys., Novosibirsk};
K. Endo, H. Hirabayashi, S. Ichii,
S. Kurokawa, T. Sato, A.Yamamoto, {\bf KEK};
M. Iwasaki {\bf Tokyo Institute of Technology}
K. Ishida,
{\bf Riken};
H. Deng, A. Disco, S.K. Dhawan, F.J.M. Farley, X. Fei, M. 
Grosse-Perdekamp
V.W. Hughes$^{\ddag}$, D. Kawall,
S.I. Redin,
{\bf Yale Univ.};
$^{\ddag}$ Spokespersons
}


%

\begin{thebibliography}{00} 

\bibitem{Dyc_90} R.~Van~Dyck,~Jr.,in {\it Quantum Electrodynamics}, T.~Kinoshita,
ed. (World Scientific, 1990), p. 322

\bibitem{Kin_95} T.~Kinoshita, IEEE Trans.Instr.Meas. {\bf 44}, 498 (1995);
Rep.Prog.Phys. {\bf 59}, 1459 (1996);
IEEE Trans. Instrum. Meas. {\bf 46}, 108 (1997)

\bibitem{QH} M. E. Cage et al., IEEE Trans. Instrum. Meas. {\bf 38}, 284 (1989)

\bibitem{ACJ} E.R. Williams et al., IEEE Trans.Instrum.Meas.{\bf38}, 233 (1989)

\bibitem{Ryd} Th. Udem et al., Phys. Rev. Lett {\bf 79}, 2646 (1997) 

\bibitem{NDBW} E. Kr\"uger et al., Metrologia {\bf 32}, 117 (1995);
IEEE Trans. Instrum. Meas. {\bf 46}, 101 (1997)

\bibitem{PDG} Particle Data Group, Phys.Rev. D{\bf 54}, 1 (1996)

\bibitem{Bos_96} M.G.Boshier et al., Comm. At. Mol. Phys. {\bf33}, 17 (1996)

\bibitem{Kin_90} T. Kinoshita and D.R. Yennie, 
in loc.cit. [1], p. 1 (1990)

\bibitem{Far_90} F.J.M.~Farley and E. Picasso, in loc.cit. [1], p.479 (1990) and
J. Bailey et al., Nucl. Phys. {\bf B 150}, 1 (1979)

\bibitem{Hug_94} BNL proposal E821 and V.W.~Hughes in {\it A Gift of Prophecy},
E.C.G. Sudarshan, ed. (World Scientific, 1994), p.222

\bibitem{Mer_90} P. Mery et al., Z.Phys. C {\bf46}, 229 (1990)

\bibitem{Lop_94} J. Lopez et al., Phys.Rev. D {\bf 49}, 366 (1994); 
G. Couture and H. Konig, Phys. Rev D {\bf 53}, 555 (1996); U. Chattopathyay 
and P. Nath, Phys. Rev. D {\bf 53}, 1648 (1996); T. Moroi Phys. Rev. D {\bf 53},
6565 (1996)

\bibitem{Ren_97} F.M. Renard et al., Phys.Lett.B {\bf 409}, 398 (1997)

\bibitem{Telegdi} V. Bargmann, L. Michel and V. Telegdi, Phys.Rev.Lett. 
{\bf 2} 435 (1959)

\bibitem{Pri_96} R.~Prigl et al., Nucl.Instr.Meth. A 304, 349 (1996)

\bibitem{Phillips_77} W.D.~Phillips et al., Metrologia {\bf 13}, 81 (1977)

\bibitem{Flegel_73} W. Flegel and F. Krienen, Nucl.Instr.Meth. {\bf 113}, 549
(1973)

\bibitem{Jun_94} K. Jungmann in {\it Atomic Physics 14}, D.J. Wineland, 
C.E. Wieman and S.J. Smith, ed. (AIP Press, 1994), p. 105

\bibitem{Prisca_97} P. Cushman, P. Nemethy et al., private communication (1997)

\bibitem{Borch_97} G. Borchert, this Euroconference

\bibitem{Yannis_97} Y. Semertzidis et al., Letter of Intent to BNL-AGS (1997)

\bibitem{Ruggiero} A. Ruggiero, BNL-Report 49529 (1993)


\end{thebibliography}
\end{document}